\documentclass[conference]{IEEEtran}
\usepackage{amssymb}
\usepackage[cmex10]{amsmath}
\usepackage{stfloats}
\usepackage{graphicx}
\usepackage{subfigure}
\usepackage{tabularx}
\usepackage{epsfig,epsf,color,balance,cite}
\usepackage{verbatim}
\usepackage{url}
\usepackage{bm}
\usepackage[a4paper,left=1.5cm,right=1.5cm, top=3cm,bottom=3.6cm,textwidth=14cm,textheight=20.3cm,
includehead,headheight=0pt,headsep=0pt]{geometry}

\newtheorem{theorem}{Theorem}

\newtheorem{lemma}{Lemma}
\usepackage{algorithm}
\usepackage{algpseudocode}
\begin{document}

% paper title
\title{Improving Wireless Physical Layer Security via D2D Communication}

\author{
\IEEEauthorblockN{Hao Xu\IEEEauthorrefmark{1},
                  Cunhua Pan\IEEEauthorrefmark{2},
                  Wei Xu\IEEEauthorrefmark{1},
                  Jianfeng Shi\IEEEauthorrefmark{1},
                  Ming Chen\IEEEauthorrefmark{1},
                  %Jiangzhou Wang\IEEEauthorrefmark{3},
and
                  Wei Heng\IEEEauthorrefmark{1}
                  }
\IEEEauthorblockA{\IEEEauthorrefmark{1}National Mobile Communications Research Laboratory, Southeast University, Nanjing 211111, China}
\IEEEauthorblockA{\IEEEauthorrefmark{2}School of Electronic Engineering and Computer Science, Queen Mary University of London, London E1 4NS, U.K.}
%\IEEEauthorblockA{\IEEEauthorrefmark{3}School of Engineering and Digital Arts, University of Kent, Canterbury, CT2 7NT, United Kingdom}
\IEEEauthorblockA{E-mail: \{xuhao2013, chenming and wheng\}@seu.edu.cn}
}

% make the title area
\maketitle

\begin{abstract}
%\boldmath
This paper investigates the physical layer security issue of a device-to-device (D2D) underlaid cellular system with a multi-antenna base station (BS) and a multi-antenna eavesdropper. To investigate the potential of D2D communication in improving network security, the conventional network without D2D users (DUs) is first considered. It is shown that the problem of maximizing the sum secrecy rate (SR) of cellular users (CUs) for this special case can be transformed to an assignment problem and optimally solved. Then, a D2D underlaid network is considered. Since the joint optimization of resource block (RB) allocation, CU-DU matching and power control is a mixed integer programming, the problem is difficult to handle. Hence, the RB assignment process is first conducted by ignoring D2D communication, and an iterative algorithm is then proposed to solve the remaining problem. Simulation results show that the sum SR of CUs can be greatly increased by D2D communication, and compared with the existing schemes, a better secrecy performance can be obtained by the proposed algorithms.

\end{abstract}

\IEEEpeerreviewmaketitle

\section{Introduction}
\label{section1}
During the past few years, wireless data traffic has grown dramatically due to the daily increasing demand on multimedia contents \cite{index2016global}. Enormous data traffic demand not only results in great pressures on the existing cellular systems, but also leads to great challenges to the transmission security since more and more confidential information is transmitted via wireless communication. Hence, enhancing the security of wireless communication has become one of the research focuses in the fifth generation (5G) network design \cite{yang2015safeguarding}. As an information-theoretic approach, physical layer security has triggered considerable research interest recently \cite{wang2012distributed, hoang2015cooperative, ding2014asymptotic, wang2015secure}.

With a great potential in improving spectral and energy efficiency, device-to-device (D2D) communication has become a promising technique \cite{xu2016channel, xu2016resource, xu2017energy}. In terms of improving the network security from the physical layer, there have been works \cite{yue2013secrecy, zhang2014radio, zhang2016joint} showing that by introducing underlaid D2D pairs to act as friendly jammers, the secrecy performance of a cellular system can be significantly increased. In \cite{yue2013secrecy}, the spectral efficiency (SE) of D2D users (DUs) was maximized with the minimum secrecy outage probability of cellular users (CUs) guaranteed. In \cite{zhang2014radio}, the problem of maximizing the system SR was considered, and the CU-DU matching problem was formulated as a weighted bipartite graph which could be solved in polynomial time. In \cite{zhang2016joint}, the sum SR of all CUs was maximized by conducting joint power and access control. However, the RB assignment and CU-DU matching were performed in a greedy manner in \cite{zhang2016joint}, which may not fully reap the RB and user diversity gains.

In this paper, we consider an uplink D2D underlaid cellular network and aim to maximize the sum SR of CUs. To improve the network security, the base station (BS) is assumed to have multiple antennas to exploit higher spatial degrees of freedom. Different from the single-input single-output (SISO) case in \cite{zhang2016joint}, where only the RB assignment strategy and the transmit power of all transmitters were optimized, in this paper, it is required to jointly optimize the variables in \cite{zhang2016joint} and the receive filters at the BS. Those variables are coupled with each other, making the optimization problem difficult to handle. Therefore, the algorithms developed in \cite{zhang2016joint} cannot be applied to solve the considered problem.

To investigate the potential of D2D communication in improving transmission security, the conventional cellular network without DUs is first considered. It is shown that the problem can be transformed to an assignment problem that can be optimally solved by applying the Hungarian algorithm \cite{kuhn1955hungarian}. The scenario with DUs is then considered. The RB assignment and CU-DU matching design make the sum SR maximization problem a mixed integer programming, which is in general difficult to solve. Hence, it is assumed that the RB assignment matrix is first obtained by ignoring D2D communication. The remaining problem is then divided into subproblems with each aiming to maximize the SR of a CU when it cooperates with a D2D pair. By adopting an iterative algorithm, a suboptimal solution to each subproblem can be obtained. Once all subproblems are solved, the CU-DU matching problem can be reformulated as an assignment problem and effectively solved. Simulation results show that the sum SR of CUs can be greatly increased by introducing DUs to serve as friendly jammers, and the proposed algorithms outperform the existing schemes remarkably in terms of sum SR since RB and user diversity gains are fully utilized.

The rest of this paper is organized as follows. In Section II, a multiuser D2D underlaid cellular system and the problem formulation are presented. In Section III, power and access control algorithms are proposed to maximize the sum SR of CUs for different cases. Numerical results are presented in Section IV before conclusions in Section V.

\section{System Model and Problem Formulation}
\subsection{System Model}
Consider an uplink communication system which consists of a BS, an eavesdropper, $M$ CUs and $N$ underlaid D2D pairs. The BS and eavesdropper are respectively equipped with $B$ and $E$ antennas, and each mobile user has one antenna. Denote the sets of CUs and D2D pairs by ${\cal{M}}$ and ${\cal{N}}$, respectively. To support multiple access for cellular communication, orthogonal frequency division multiplexing (OFDM) is adopted \cite{zhu2012radio}, and there is a set ${\cal{K}}$ of $K$ orthogonal RBs. For brevity, a fully loaded cellular network scenario is assumed, i.e., $K=M$. As in \cite{yue2013secrecy, zhang2014radio, zhang2016joint}, assume that each CU cooperates with at most one D2D pair and each D2D pair reuses the RB of no more than one CU. Moreover, it is also assumed that global channel state information (CSI) is available at the BS.

Define $\alpha_m^k$ to indicate whether CU $m$ uses RB $k$ or not, i.e., if CU $m$ uses RB $k$, $\alpha_{m}^k=1$, otherwise, $\alpha_{m}^k=0$. Introduce variable $\theta_m^n$ to indicate whether CU $m$ cooperates with D2D pair $n$ or not, i.e., if CU $m$ shares its RB with D2D pair $n$, $\theta_m^n=1$, otherwise, $\theta_m^n=0$. For signal detection, the BS is assumed to adopt minimum mean-square error (MMSE) receiver $\bm w_m \in {\mathbb C}^{B\times 1}$ to detect the data symbol of CU $m$. Then, if CU $m$ cooperates with D2D pair $n$ on RB $k$, i.e., $\alpha_m^k=\theta_m^n=1$, the post-processing signal-to-interference-plus-noise ratio (SINR) from CU $m$ to the BS is given by
\begin{equation}
{\eta}_{m,n}^k = \frac{ p_m \left|\bm w_m^H \bm g_{mb}^k\right|^2}{q_n \left|\bm w_m^H \bm h_{nb}^k\right|^2+N_0 \left\|\bm w_m \right\|^2},
\label{SINR_CU}
\end{equation}
where $p_m$ denotes the transmit power of CU $m$, and ${\bm g}_{mb}^k \in {\mathbb C}^{B\times1}$ represents the channel vector from CU $m$ to the BS on RB $k$. $q_n$ and ${\bm h}_{nb}^k$ are similarly defined for D2D transmitter (D2D-Tx) $n$, and $N_0$ is the noise power.

Since calculating MMSE receivers requires conducting the inversion of matrices, which usually involves high computational complexity, the eavesdropper is assumed to simply employ maximal-ratio combining (MRC) for signal detection. Then, if $\alpha_m^k=\theta_m^n=1$, the post-processing SINR from CU $m$ to the eavesdropper on RB $k$ is
\begin{equation}
{\gamma}_{m,n}^k = \frac{ p_m \left\|\bm g_{me}^k\right\|^4}{q_n \left|\left(\bm g_{me}^k\right)^H \bm h_{ne}^k\right|^2+N_0 \left\|\bm g_{me}^k\right\|^2},
\label{SINR_eav}
\end{equation}
where $\bm g_{me}^k, \bm h_{ne}^k \in {\mathbb C}^{E\times1}$ denote the channel vectors from CU $m$ and D2D-Tx $n$ to the eavesdropper on RB $k$, respectively.

Thus, if $\alpha_m^k=\theta_m^n=1$,  the SR of CU $m$ on RB $k$ can be calculated as \cite{barros2006secrecy}
\begin{equation}
C_{m,n}^k = \frac{V}{K} \left[\log_2\left(1+{\eta}_{m,n}^k\right) - \log_2\left(1+{\gamma}_{m,n}^k\right)\right]^+,
\label{secrecy_capacity}
\end{equation}
where $V$ is the total bandwidth and $[\cdot]^+ \triangleq \max (\cdot,0)$.
\subsection{Problem Formulation}
This paper aims to maximize the sum SR of all CUs with the help of D2D communication. The problem can be mathematically formulated as
{\setlength\arraycolsep{2pt}%用于缩短等号两边的空格
\begin{subequations}
\begin{align}
\mathop {\max }\limits_{{\bm A}, {\bm \Theta}, {\bm p}, {\bm q}, {\bm W}} \quad & \sum\limits_{m \in {\cal M}} \sum\limits_{n \in {\cal N}} \sum\limits_{k \in {\cal K}} \alpha_m^k \theta_m^n C_{m,n}^k \label{optimize_problem_a}\\
\text{s.t.} \quad\;\, &  0 \leq p_m \leq P_m, \quad\forall m \in {\cal M}, \label{optimize_problem_b}\\
&  0 \leq q_n \leq Q_n, \quad\forall n \in {\cal N}, \label{optimize_problem_c}\\
& \alpha_{m}^k \in \left\{0,1\right\} , \quad\forall m \in {\cal M}, k \in {\cal K},\label{optimize_problem_d}\\
& \sum\limits_{k \in {\cal K}} \alpha_m^k =1, \quad\forall m \in {\cal M},\label{optimize_problem_e}\\
& \sum\limits_{m \in {\cal M}} \alpha_m^k=1, \quad\forall k \in {\cal K},\label{optimize_problem_f}\\
& \theta_m^n \in \left\{0,1\right\} , \quad\forall m \in {\cal M}, n \in {\cal N},\label{optimize_problem_g}\\
& \sum\limits_{n \in {\cal N}} \theta_m^n \leq 1, \quad\forall m \in {\cal M},\label{optimize_problem_h}\\
& \sum\limits_{m \in {\cal M}} \theta_m^n\leq 1, \quad\forall n \in {\cal N},\label{optimize_problem_i}
\end{align}
\label{optimize_problem}
\end{subequations}}
\!\!\!\!\!where ${\bm A}=[\alpha_m^k]_{M\times K}$, $\bm \Theta=[\theta_m^n]_{M\times N}$, ${\bm p}=(p_1, \cdots, p_M)^T$, ${\bm q}=(q_1, \cdots, q_N)^T$ and ${\bm W}=[\bm w_1, \cdots, \bm w_M]$. $P_m$ and $Q_n$ are the maximum transmit powers of CU $m$ and D2D-Tx $n$. (\ref{optimize_problem_e}) and (\ref{optimize_problem_f}) indicate that a CU should be allocated an RB and an RB can be used by only one CU. (\ref{optimize_problem_h}) and (\ref{optimize_problem_i}) correspond to the RB reuse strategy which has been stated above.

\section{Physical Layer Security Based Power and Access Control}
\label{section3}
In this section, we first consider a simplified special case of (\ref{optimize_problem}), i.e., the conventional cellular communication without DUs, to characterize the potential of D2D communication in improving network security. It will be shown that the corresponding problem can be optimally solved. Then, we focus on solving problem (\ref{optimize_problem}). Beforehand, we first give Lemma \ref{theorem_1}, which is useful for the following analysis.
\begin{lemma}
Assume that ${\bm A}^*$, ${\bm p}^*$, ${\bm q}^*$ and ${\bm W}^*$ are the optimal solutions to (\ref{optimize_problem}), and denote the RB assigned to CU $m$ by $k_m$, i.e., $\alpha_m^{k_m*}=1$. Then, the transmit power of CU $m$ satisfies
\begin{equation}
p_m^* = \left\{
\begin{array}{ll}
P_m,&  \frac{ \left|\bm w_m^{*H} \bm g_{mb}^{k_m}\right|^2}{\left\|\bm g_{me}^{k_m}\right\|^4 } > \frac{ q_n \left|\bm w_m^{*H} \bm h_{nb}^{k_m}\right|^2+N_0 \left\|\bm w_m^* \right\|^2}{q_n \left|\left(\bm g_{me}^{k_m}\right)^H \bm h_{ne}^{k_m}\right|^2+N_0 \left\|\bm g_{me}^{k_m}\right\|^2}\\
0,&  {\text {otherwise}}\\
\end{array} \right..
\label{p_value}
\end{equation}
\label{theorem_1}
\end{lemma}
\itshape \textbf{Proof:}  \upshape
See Appendix A.
\hfill $\Box$

Lemma \ref{theorem_1} indicates that under the optimal condition, each CU either keeps inactive (i.e., the corresponding transmit power is 0) or transmits its signal with the maximum power.
\subsection{Without D2D communication}
\label{III-A}
When $N=0$, problem (\ref{optimize_problem}) can be simplified as
{\setlength\arraycolsep{2pt}%用于缩短等号两边的空格
\begin{eqnarray}
\mathop {\max }\limits_{{\bm A}, {\bm p}, {\bm W}} && \sum\limits_{m \in {\cal M}} \sum\limits_{k \in {\cal K}} \alpha_m^k C_m^k \label{optimize_problem1_a}\nonumber\\
\text{s.t.} \;\, && (\text{\ref{optimize_problem_b}}),(\text{\ref{optimize_problem_d}})\sim (\text{\ref{optimize_problem_f}}),
\label{optimize_problem1}
\end{eqnarray}}
\!\!where $C_m^k$ is a simplified version of $C_{m,n}^k$ and is given by
\begin{equation}
C_{m}^k \!=\! \frac{V}{K}\!\! \left[\!\log_2\!\!\left(\!\!1\!+\!\frac{ p_m \left|\bm w_m^H \bm g_{mb}^k\right|^2}{N_0 \left\|\bm w_m \right\|^2}\!\right) \!-\! \log_2\!\!\left(\!\!1\!+\!\frac{ p_m \left\|\bm g_{me}^k\right\|^2}{N_0}\!\right)\!\!\right]^+\!\!.
\label{secrecy_capacity1}
\end{equation}

To the best of the authors' knowledge, the power and access control problem (\ref{optimize_problem1}) has not been studied. From Lemma \ref{theorem_1}, it is known that when $\alpha_m^k = 1$, CU $m$ either keeps inactive or transmits in the maximum power on RB $k$. If $p_m=P_m$, the MMSE receiver at the BS on RB $k$ for detecting the signal of CU $m$ can be directly obtained as
\begin{equation}
{\tilde {\bm w}}_m^k = \sqrt{P_m} \left[ P_m \bm g_{mb}^k \left(\bm g_{mb}^k\right)^H + N_0 \bm I_B\right]^{-1} \bm g_{mb}^k.
\label{MMSE_receiver}
\end{equation}
Then, the SR of CU $m$ on RB $k$ is formulated as
{\setlength\arraycolsep{2pt}%用于缩短等号两边的空格
\begin{eqnarray}
\phi_{m}^k &=& \frac{V}{K} \left[\log_2\left(1+\frac{P_m \left|\left({\tilde {\bm w}}_m^k\right)^H \bm g_{mb}^k\right|^2}{N_0 \left\|{\tilde {\bm w}}_m^k \right\|^2}\right)\right.\nonumber\\
&&-\left.\log_2\left(1+\frac{P_m \left\|\bm g_{me}^k\right\|^2}{N_0}\right)\right]^+.
\label{secrecy_capacity2}
\end{eqnarray}}
 \!\!\!The optimal channel allocation matrix $\bm A$ is thus determined by the following problem
{\setlength\arraycolsep{2pt}%用于缩短等号两边的空格
\begin{eqnarray}
\mathop {\max }\limits_{\bm A} && \sum\limits_{m \in {\cal M}} \sum\limits_{k \in {\cal K}} \alpha_m^k \phi_m^k \label{optimize_problem2_a}\nonumber\\
\text{s.t.} \;\, && (\text{\ref{optimize_problem_d}})\sim (\text{\ref{optimize_problem_f}}),
\label{optimize_problem2}
\end{eqnarray}}
\!\!\!which aims to maximize the sum SR of CUs by finding the optimal ${\bm A}$ with no more than two 1's in the same row or column, and is thus an assignment problem. Obviously, problem (\ref{optimize_problem2}) is equivalent to (\ref{optimize_problem1}) in terms of sum SR of CUs, and can be optimally solved by using the Hungarian algorithm \cite{kuhn1955hungarian}. Once the optimal $\bm A$ is obtained, we have $\bm w_m=\sum\limits_{k \in {\cal K}} \alpha_m^k {\tilde {\bm w}}_m^k$ and the optimal $\bm p$ from Lemma \ref{theorem_1}.

Thus, the power and access control algorithm for the case without D2D communication (referred to `PAC-NoD2D') can be summarized in Algorithm \ref{PAC-NoD2D}. For brevity, please refer to \cite{kuhn1955hungarian} for detailed description of the Hungarian algorithm.
\begin{algorithm}[h]
\begin{algorithmic}[1]
\caption{PAC-NoD2D}
\label{PAC-NoD2D}
\State \text {Initialization:} Calculate matrix $\Phi=[-\phi_m^k]_{M\times K}$ using (\ref{secrecy_capacity2}) and (\ref{MMSE_receiver}).
\State  Obtain the optimal ${\bm A}$ by using the Hungarian algorithm based on \cite{kuhn1955hungarian}.
\State  $\bm w_m=\sum\limits_{k \in {\cal K}} \alpha_m^k {\tilde {\bm w}}_m^k$. Obtain the optimal $\bm p$ based on Lemma \ref{theorem_1}.
\end{algorithmic}
\end{algorithm}

According to Lemma \ref{theorem_1}, an active CU always transmits its signal using the maximum power. Hence, in the following corollary, the effect of the maximum transmit power of a CU on its corresponding SR is investigated.
\begin{lemma}
For any $m \in {\cal M}$ and $k \in {\cal K}$, whether $\phi_m^k$ is positive or not depends only on the relation between $\left\|\bm g_{mb}^k\right\|$ and $\left\|\bm g_{me}^k\right\|$. When $\left\|\bm g_{mb}^k\right\|>\left\|\bm g_{me}^k\right\|$, $\phi_m^k$ increases with $P_m$, and in the high $P_m$ regime, i.e., high signal-to-noise ratio (SNR) case, $\phi_m^k$ becomes asymptotically invariant with $P_m$.
\label{lemma1}
\end{lemma}
\itshape \textbf{Proof:}  \upshape
See Appendix \ref{Appendix_B}.
\hfill $\Box$

\subsection{With D2D Communication}
\label{Single-reuse}
When D2D communication is adopted to improve the system security, we need to solve problem (\ref{optimize_problem}). Since each element of matrices ${\bm A}$ and $\bm \Theta$ is binary valued, problem (\ref{optimize_problem}) is a mixed integer programming, which is usually intractable. To make it tractable, assume that ${\bm A}$ is first obtained based on Algorithm \ref{PAC-NoD2D} by ignoring D2D communication. Then, it is necessary to solve the CU-DU matching and power control problem as follows
{\setlength\arraycolsep{2pt}%用于缩短等号两边的空格
\begin{eqnarray}
\mathop {\max }\limits_{\bm \Theta, {\bm p}, {\bm q}, {\bm W}} \quad && \sum\limits_{m \in {\cal M}} \sum\limits_{n \in {\cal N}} \theta_m^n C_{m,n}^{k_m} \label{optimize_problem4_a}\nonumber\\
\text{s.t.} \quad\;\,\, &&  (\text{\ref{optimize_problem_b}}), (\text{\ref{optimize_problem_c}}), (\text{\ref{optimize_problem_g}})\sim(\text{\ref{optimize_problem_i}}).
\label{optimize_problem4}
\end{eqnarray}}
\!\!From Lemma \ref{theorem_1}, it is known that in the optimal case, each CU either keeps inactive or transmits its signal with the maximum power. Denoting
{\setlength\arraycolsep{2pt}%用于缩短等号两边的空格
\begin{eqnarray}
&&\chi_{m,n}^{k_m} =\log_2\left(1+\frac{ P_m \left|\bm w_m^H \bm g_{mb}^{k_m}\right|^2}{q_n \left|\bm w_m^H \bm h_{nb}^{k_m}\right|^2+N_0 \left\|\bm w_m \right\|^2}\right) \nonumber\\
&&\quad- \log_2\!\left(\!1+\frac{ P_m \left\|\bm g_{me}^{k_m}\right\|^4}{q_n \left|\left(\bm g_{me}^{k_m}\right)^H \bm h_{ne}^{k_m}\right|^2+N_0 \left\|\bm g_{me}^{k_m}\right\|^2}\right),\quad\quad
\label{secrecy_capacity4}
\end{eqnarray}}
\!\!\!then, if $\theta_m^n=1$, the maximum SR of CU $m$ can be obtained by solving the following problem
{\setlength\arraycolsep{2pt}%用于缩短等号两边的空格
\begin{subequations}
\begin{align}
\mathop {\max }\limits_{q_n, {\bm w}_m} \quad & \chi_{m,n}^{k_m} \label{optimize_problem5_a}\\
\text{s.t.} \quad\;\, & 0\leq q_n \leq Q_n.\label{optimize_problem5_b}
\end{align}
\label{optimize_problem5}
\end{subequations}}

When the BS and the eavesdropper both have one antenna, i.e., $B=1$ and $E=1$, the optimal $q_n$ can be obtained as shown in \cite{zhang2016joint}. When $B>1$, problem (\ref{optimize_problem5}) becomes more complex since $q_n$ and ${\bm w}_m$ are coupled, and the fractional expression of SINRs as well as the $\log_2 (\cdot)$ operation makes problem (\ref{optimize_problem5}) nonconvex. To deal with the SINR in fractional form, the relationship between the MMSE and the pre-processing SINR is usually applied in a downlink cellular system \cite{xu2015robust}, \cite{christensen2008weighted}. As for the considered D2D underlaid uplink cellular system, a similar relationship as presented in Lemma \ref{lemma2} also holds. Since Lemma \ref{lemma2} can be analogously verified as that in \cite{christensen2008weighted}, we omit the proof process for brevity.

\begin{lemma}
If $\alpha_m^k=\theta_m^n=1$, and $\bm w_m$ is an MMSE receiver, the following relationship between the MMSE and the post-processing SINR of cellular link $m$ holds
\begin{equation}
{\text {MMSE}}_m=\frac{1}{1+\eta_{m,n}^k},
\label{MMSE_SINR}
\end{equation}
where ${\text {MMSE}}_m$ denotes the MMSE of cellular link $m$.
\label{lemma2}
\end{lemma}

When $\theta_m^n=1$, the MMSE of CU $m$ is given by
\begin{equation}
{\text {MMSE}}_m = \left|\sqrt{P_m} \bm w_m^H \bm g_{mb}^{k_m}\!-\!1\right|^2 \!+ q_n \!\left|\bm w_m^H \bm h_{nb}^{k_m}\right|^2\!+\!N_0 \left\|\bm w_m \right\|^2.
\label{MMSE_CU1}
\end{equation}
Since $\bm w_m$ is an MMSE receiver, according to Lemma \ref{lemma2}, $\chi_{m,n}^{k_m}$ is equivalent to
{\setlength\arraycolsep{2pt}%用于缩短等号两边的空格
\begin{eqnarray}
\!\!\!&&{\tilde \chi}_{m,n}^{k_m} =\nonumber\\
\!\!\!&&-\log_2\!\left(\left|\sqrt{P_m} \bm w_m^H \bm g_{mb}^{k_m}\!-\!1\right|^2 \!\!+\! q_n \left|\bm w_m^H \bm h_{nb}^{k_m}\right|^2\!\!+\!N_0 \left\|\bm w_m \right\|^2\right) \nonumber\\
\!\!\!&&- \log_2\!\left(\!1+\frac{ P_m \left\|\bm g_{me}^{k_m}\right\|^4}{q_n \left|\left(\bm g_{me}^{k_m}\right)^H \bm h_{ne}^{k_m}\right|^2+N_0 \left\|\bm g_{me}^{k_m}\right\|^2}\right).\quad\quad
\label{secrecy_capacity5}
\end{eqnarray}}
\!\!\!Then, problem (\ref{optimize_problem5}) can be transformed to
{\setlength\arraycolsep{2pt}%用于缩短等号两边的空格
\begin{subequations}
\begin{align}
\mathop {\max }\limits_{q_n, {\bm w}_m} \quad & {\tilde \chi}_{m,n}^{k_m}\\
\text{s.t.} \quad\;\, & 0\leq q_n \leq Q_n.
\end{align}
\label{optimize_problem51}
\end{subequations}}
\!\!\!\!Obviously, problem (\ref{optimize_problem51}) is equivalent to problem (\ref{optimize_problem5}) and can be effectively solved by iteratively optimizing ${\bm w}_m$ and $q_n$. For fixed $q_n$, the optimal MMSE receive filter for detecting the signal of CU $m$ can be directly written as
\begin{equation}
{\bm w}_m^* \!=\! \sqrt{P_m} \!\left[\! P_m {\bm g}_{mb}^{k_m} \!\left(\! {\bm g}_{mb}^{k_m}\!\right)^{\!H} \!\!\!+\! q_n {\bm h}_{nb}^{k_m} \!\left(\! {\bm h}_{nb}^{k_m}\!\right)^{\!H} \!\!\!+\! N_0 \bm I_B\!\right]^{\!-1} \!\!\!{\bm g}_{mb}^{k_m}.
\label{MMSE1}
\end{equation}
When ${\bm w}_m$ is determined, the optimal $q_n$ can be obtained based on the following theorem.
\begin{theorem}
When $\theta_m^n=1$ and ${\bm w}_m$ is determined, the optimal transmit power of D2D-Tx $n$ is given by
\begin{equation}
q_n^* = \left\{
\begin{array}{ll}
{\tilde a}_m^n,&  \Delta_m^n > 0 \,{\text {and}}\, 0<{\tilde a}_m^n < Q_n\\
Q_n,&  \Delta_m^n > 0 \,{\text {and}}\, {\tilde a}_m^n \geq Q_n\\
0,&  {\text {otherwise}}\\
\end{array} \right.,
\label{q_value}
\end{equation}
where ${\tilde a}_m^n$ and $\Delta_m^n$ are respectively defined in (\ref{zero_points}) and (\ref{discriminant}).
\label{theorem_3}
\end{theorem}
\itshape \textbf{Proof:}  \upshape
See Appendix \ref{Appendix_C}.
\hfill $\Box$

By iteratively updating ${\bm w}_m$ and $q_n$ based on (\ref{MMSE1}) and Theorem \ref{theorem_3}, a suboptimal ${\tilde \chi}_{m,n}^{k_m}$ can be obtained. Then, denote $\psi_{m,n}^{k_m}=\frac{V}{K}[{\tilde \chi}_{m,n}^{k_m}]^+$, the CU-DU matching problem becomes
{\setlength\arraycolsep{2pt}%用于缩短等号两边的空格
\begin{eqnarray}
\mathop {\max }\limits_{\bm \Theta} \quad && \sum\limits_{m \in {\cal M}} \sum\limits_{n \in {\cal N}} \theta_m^n \psi_{m,n}^{k_m} \label{optimize_problem6_a}\nonumber\\
\text{s.t.} \quad\; && (\text{\ref{optimize_problem_g}})\sim(\text{\ref{optimize_problem_i}}).
\label{optimize_problem6}
\end{eqnarray}}
\!\!Obviously, problems (\ref{optimize_problem4}) and (\ref{optimize_problem6}) are equivalent in terms of sum SR of CUs. Denote $\Psi= \left[-\psi_{m,n}^{k_m} \right]_{M\times N}$. When $M=N$, $\Psi$ is a square matrix and constraints (\ref{optimize_problem_h}), (\ref{optimize_problem_i}) will hold with equality. In this case, problem (\ref{optimize_problem6}) is an assignment problem and can be optimally solved by employing the Hungarian algorithm. Once $\bm \Theta$ is determined, ${\bm q}$ and ${\bm W}$ can be directly optimized from previously obtained solution of problem (\ref{optimize_problem5}), and ${\bm p}$ can then be obtained based on Lemma \ref{theorem_1}.

When $M<N$, (\ref{optimize_problem6}) can be cast to an assignment problem by first calculating $\Psi$ based on the above analysis and then adding an $(N-M)\times N$ zero matrix to the bottom of $\Psi$ to obtain a new $N\times N$ dimensional matrix $\tilde {\Psi}$. Similarly, an $M\times (M-N)$ zero matrix can be added to the right side of $\Psi$ to obtain a new $M\times M$ dimensional matrix $\tilde {\Psi}$ to solve problem (\ref{optimize_problem6}) when $M>N$. Note that in this case, since the number of D2D pairs is smaller than that of CUs, only $N$ CUs can get help from DUs to improve their secrecy performance, and the rest $M-N$ CUs have to transmit in the conventional cellular communication mode.

Based on the above analysis, the power and access control algorithm for the case with D2D communication (referred to `PAC-D2D') can be summarized in Algorithm \ref{PAC-SR}.

\begin{algorithm}[h]
\begin{algorithmic}[1]
\caption{PAC-D2D}
\label{PAC-SR}
\State Obtain ${\bm A}$ based on Algorithm \ref{PAC-NoD2D} by ignoring D2D communication. Set $\bm q =\bm 0$.
\For {$m=1,\cdots, M$}
\For {$n=1,\cdots, N$}
\Repeat
\State Calculate $\bm w_m$ based on (\ref{MMSE1}),
\State Calculate $q_n$ based on Theorem \ref{theorem_3},
\Until convergence
\State Calculate ${\tilde \chi}_{m,n}^{k_m}$ from (\ref{secrecy_capacity5}). $\psi_{m,n}^{k_m}=\frac{V}{K}[{\tilde \chi}_{m,n}^{k_m}]^+$.
\EndFor
\EndFor
\State $\Psi= \left[-\psi_{m,n}^{k_m} \right]_{M\times N}$.
\State If $M=N$, solve problem (\ref{optimize_problem6}) and obtain $\bm \Theta$ by applying the Hungarian algorithm. Otherwise, if $M<N$ ($M>N$), add an $(N-M)\times N$ zero matrix to the bottom of $\Psi$ (an $M\times (M-N)$ zero matrix on the right side of $\Psi$) and obtain a new square matrix $\tilde {\Psi}$. Solve the resulted assignment problem and obtain $\bm \Theta$ by applying the Hungarian algorithm.
\State Obtain ${\bm p}$ based on Lemma \ref{theorem_1}.
\label{algorithm2}
\end{algorithmic}
\end{algorithm}

\subsection{Convergence Analysis}
To verify the convergence of the proposed PAC-D2D algorithm, we only need to prove that the iteration of updating $\bm w_m$ and $q_n$ converges for any $m \in {\cal M}$ and $n \in {\cal N}$. In each iteration, the optimal $\bm w_m$ is first obtained based on (\ref{MMSE1}) with fixed $q_n$. Then, for determined $\bm w_m$, the PAC-D2D algorithm outputs an optimal $q_n$ based on Theorem \ref{theorem_3}. As a result, ${\tilde \chi}_{m,n}^{k_m}$ always increases after each iteration. Since ${\tilde \chi}_{m,n}^{k_m}$ is upper bounded, the iterative updating of $\bm w_m$ and $q_n$ converges, which guarantees the convergence of the PAC-D2D algorithm.

\subsection{Complexity Analysis}
In this subsection, the computational complexity of Algorithm \ref{PAC-NoD2D} and Algorithm \ref{PAC-SR} is analyzed with order notation.

The complexity of the PAC-NoD2D algorithm mainly lies in solving problem (\ref{optimize_problem2}) by using the Hungarian algorithm, which has a complexity of ${\cal O}(M^4)$. As for the PAC-D2D algorithm, the complexity mainly lies in calculating $\bm w_m$, whose complexity is dominated by the inversion process in (\ref{MMSE1}). According to \cite{he2014coordinated}, inversion of an $B\times B$ Hermite matrix requires $\frac{4}{3}B^3$ floating point operations. Therefore, calculating (\ref{MMSE1}) involves a complexity of ${\cal O}\left(\frac{4}{3}B^3\right)$. Denote the number of iterations for updating $\bm w_m$ and $q_n$ by $I$. Since the iteration is carried out $MN$ times, the total complexity of the PAC-D2D algorithm is ${\cal O}\left(\frac{4}{3}B^3 MN I\right)$. Simulation results show that the PAC-D2D algorithm converges rapidly, so the complexity is low and acceptable.

\section{Simulation Results}
\begin{table}\centering
\label{Simu_Para}
\caption{Simulation Parameters}
\vspace{-0.8em}
\begin{tabular}{|l|l|}
\hline Radius of the cell & 500 m \\
\hline Maximum distance of a D2D pair $D_{\text{max}}$ & 50\\
\hline Additive noise power $N_0$ & -100 dBm\\
\hline Path loss exponent & 3.7\\
\hline Standard deviation of log-normal shadowing fading& 8 dB\\
\hline
\end{tabular}
\end{table}
In this section, simulation results are presented to evaluate the performance of the proposed algorithms. Consider a single-cell network where the BS locates at the center and the eavesdropper as well as all mobile users are randomly distributed. The distance between a D2D-Tx and its associated receiver is uniformly distributed in $[0\,\text{m}, D_{\text{max}}\,\text{m}]$. For brevity, unit total bandwidth and equal maximum power constraint are assumed, i.e., $V=1$ Hz and $P_m= Q_n=P, \forall m\in {\cal M}, n\in {\cal N}$. The other parameters are summarized in Table I unless otherwise specified. All simulation results are obtained by averaging over $1000$ random topologies. To investigate the performance of the proposed algorithms, the PAC-NOD2D algorithm is compared with the scheme (labeled as `NoD2D-random') which randomly assigns RBs to CUs, and the PAC-D2D algorithm is compared with the power and access control algorithm proposed in \cite{zhang2016joint} (labeled as `PAC in \cite{zhang2016joint}').

\label{section5}
\begin{figure}
  \centering
  \includegraphics[scale=0.5]{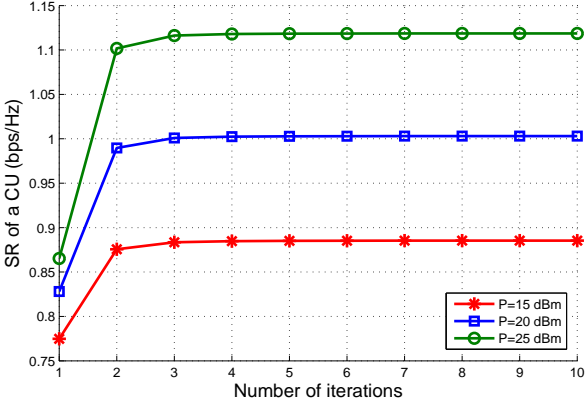}
  \caption{Convergence behaviors of the proposed PAC-D2D algorithm with $M=K=6$, $N=10$ and $B=E=4$.}
  \label{Fig.1}
\end{figure}

Fig. \ref{Fig.1} illustrates the convergence behavior of iteratively updating $\bm w_m$ and $q_n$ for the proposed PAC-D2D algorithm. The vertical axis represents the SR of CU $m$ when it cooperates with D2D pair $n$. It can be seen that the SR of CU $m$ increases monotonically during the iterative process and converges rapidly for all considered configurations. Two to three iterations are sufficient for PAC-D2D to converge.

\begin{figure}
  \centering
  \includegraphics[scale=0.5]{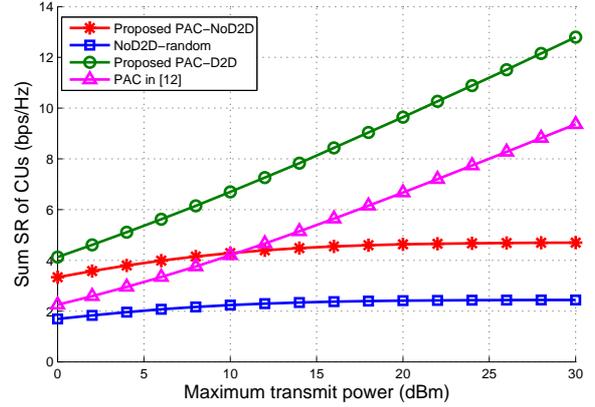}
  \caption{Sum SR of CUs versus the maximum transmit power with $M=K=6$, $N=10$ and $B=E=4$.}
  \label{Fig.2}
\end{figure}

The sum SR of CUs versus the maximum power is depicted in Fig. \ref{Fig.2}, from which several observations can be made. First, as expected, the sum SR of CUs increases with $P$ for all considered cases. Second, since RBs are assigned to CUs optimally, the sum SR of CUs obtained by the proposed PAC-NoD2D algorithm is much larger than that of NoD2D-random. For both PAC-NoD2D and NoD2D-random algorithms, as concluded in Lemma \ref{lemma1}, the sum SR of CUs always grows with $P$, and the increase becomes less obvious as $P$ grows large. Moreover, Fig. \ref{Fig.2} also shows that the cellular transmission security can be significantly improved by D2D communication, and the proposed PAC-D2D algorithm outperforms PAC in \cite{zhang2016joint} greatly in terms of cellular sum SR.

\begin{figure}
  \centering
  \includegraphics[scale=0.5]{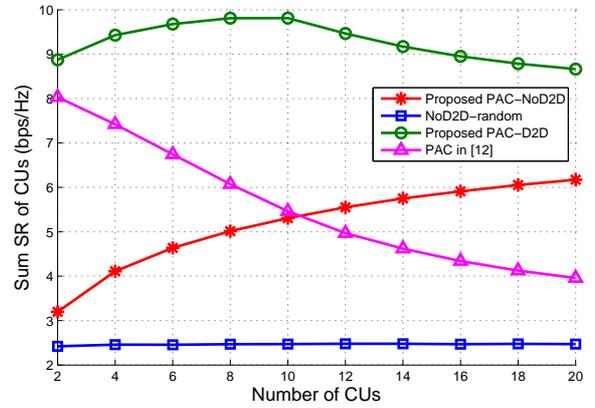}
  \caption{Sum SR of CUs versus the number of CUs with $N=10$, $B=E=4$ and $P=20$ dBm.}
  \label{Fig.3}
\end{figure}

In Fig. \ref{Fig.3}, the sum SR of CUs versus the number of CUs is depicted. It can be seen that as $M$ increases, the sum SR of CUs remains unchanged for NoD2D-random, while grows for PAC-NoD2D. This is because RBs are randomly assigned to CUs by NoD2D-random, while optimally allocated by PAC-NoD2D, which can fully reap the  RB diversity gains. For the case with DUs, it can be found that as $M$ grows, the sum SR of CUs decreases for PAC in \cite{zhang2016joint}, while first increases and then reduces for the proposed PAC-D2D algorithm. This is because the RB assignment process and CU-DU matching are performed in a greedy manner by PAC in \cite{zhang2016joint}. For fixed $N$, a high-numbered CU always has less choices in selecting its cooperating D2D pair than a low-numbered CU. Hence, the sum SR of CUs decreases with $M$. As for the proposed PAC-D2D algorithm, the RB assignment process and CU-DU matching are optimally performed via solving the corresponding assignment problems, so the sum SR of CUs increases with $M$ when $M\leq N$. After that, the sum SR of CUs reduces since $M-N$ CUs will be unable to cooperate with a D2D pair.

\begin{figure}
  \centering
  \includegraphics[scale=0.5]{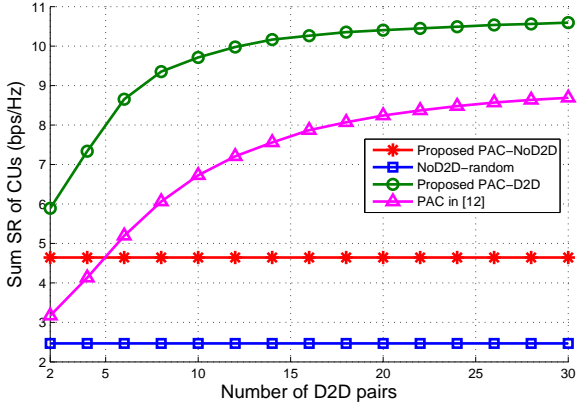}
  \caption{Sum SR of CUs versus the number of D2D pairs with $M=K=6$, $B=E=4$ and $P=20$ dBm.}
  \label{Fig.4}
\end{figure}

In Fig. \ref{Fig.4}, the effect of the number of D2D pairs is depicted. Obviously, $N$ will not affect the curves corresponding to the NoD2D-random and PAC-NoD2D algorithms. For the case with DUs, the sum SR of CUs increases with $N$, and the proposed PAC-D2D algorithm outperforms PAC in \cite{zhang2016joint} greatly. In particular, when $N=10$, in contrast to PAC in \cite{zhang2016joint}, the sum SR of CUs can be increased by over 40\% via adopting the proposed PAC-D2D algorithm.
\section{Conclusions}
\label{section6}
This paper has studied the transmission secrecy of a D2D underlaid cellular network. To investigate the potential of D2D communication, the special case without DUs was first considered. It was shown that the sum SR maximization problem for this case could be optimally solved. Then, the case with DUs was considered and an iterative algorithm was proposed to solve the corresponding problem. Simulation results showed that the sum SR of CUs could be significantly increased by introducing D2D communication, and the proposed algorithms outperformed the existing schemes greatly in terms of sum SR.
\appendices
\section{Proof of Lemma \ref{theorem_1}}
\label{Appendix_A}
If $\frac{ \left|\bm w_m^{*H} \bm g_{mb}^{k_m}\right|^2}{\left\|\bm g_{me}^{k_m}\right\|^4 } \leq \frac{ q_n \left|\bm w_m^{*H} \bm h_{nb}^{k_m}\right|^2+N_0 \left\|\bm w_m^* \right\|^2}{q_n \left|\left(\bm g_{me}^{k_m}\right)^H \bm h_{ne}^{k_m}\right|^2+N_0 \left\|\bm g_{me}^{k_m}\right\|^2}$, from equations (\ref{SINR_CU}), (\ref{SINR_eav}) and (\ref{secrecy_capacity}), it is known that a positive SR cannot be obtained on this RB $k_m$. Hence $p_m^*=0$. Otherwise, if $\frac{ \left|\bm w_m^{*H} \bm g_{mb}^{k_m}\right|^2}{\left\|\bm g_{me}^{k_m}\right\|^4 } > \frac{ q_n \left|\bm w_m^{*H} \bm h_{nb}^{k_m}\right|^2+N_0 \left\|\bm w_m^* \right\|^2}{q_n \left|\left(\bm g_{me}^{k_m}\right)^H \bm h_{ne}^{k_m}\right|^2+N_0 \left\|\bm g_{me}^{k_m}\right\|^2}$, under the help of D2D communication, the BS possesses better channel conditions of CU $m$ than the eavesdropper on RB $k_m$. In this case, it can be found from (\ref{secrecy_capacity}) that $C_{m,n}^{k_m}$ increases with $p_m$. Hence, the maximum SR can be obtained by letting CU $m$ transmit its signal on RB $k_m$ using the maximum power, i.e., $p_m^*=P_m$. The proof is completed.

\section{Proof of Lemma \ref{lemma1}}
\label{Appendix_B}
For the case without D2D communication, $\eta_{m,n}^k$ can be simplified as $\eta_m^k$ by omitting subscript $n$. When MMSE receivers are adopted for signal detection, it can be readily verified that $\eta_m^k$ satisfies
\begin{equation}
\eta_m^k = \frac{ p_m \left|\bm w_m^H \bm g_{mb}^k\right|^2}{N_0 \left\|\bm w_m \right\|^2}=\frac{ p_m \left\|\bm g_{mb}^k\right\|^2}{N_0}.
\end{equation}
Hence, (\ref{secrecy_capacity2}) can be rewritten as
{\setlength\arraycolsep{2pt}% 用于缩短等号两边的空格
\begin{eqnarray}
\phi_{m}^k(P_m) &=& \frac{V}{K}\left[\log_2\left(1+\frac{ P_m \left\|\bm g_{mb}^k\right\|^2}{N_0}\right)\right.\nonumber\\
&&- \left.\log_2\left(1+\frac{P_m \left\|\bm g_{me}^k\right\|^2}{N_0}\right)\right]^+.
\label{secrecy_capacity21}
\end{eqnarray}}
\!\!As a function of $P_m$, the first-order derivative of the term inside $[\cdot]^+$ in (\ref{secrecy_capacity21}) can be calculated as
\begin{equation}
\frac{N_0 \left(\left\|\bm g_{mb}^k\right\|^2-\left\|\bm g_{me}^k\right\|^2\right)}{\left(N_0+P_m \left\|\bm g_{mb}^k\right\|^2\right)\left(N_0+P_m \left\|\bm g_{me}^k\right\|^2\right)\ln 2}.
\label{derivative0}
\end{equation}
When $\left\|\bm g_{mb}^k\right\|\leq \left\|\bm g_{me}^k\right\|$, (\ref{derivative0}) is nonpositive, indicating that the term inside $[\cdot]^+$ of (\ref{secrecy_capacity21}) decreases with $P_m$. So $\phi_{m}^k=\phi_{m}^k(0)=0$. When $\left\|\bm g_{mb}^k\right\|>\left\|\bm g_{me}^k\right\|$, $\phi_{m}^k$ increases with $P_m$ and is positive. Therefore, whether $\phi_{m}^k$ is positive or not is only determined by the relationship between $\left\|\bm g_{mb}^k\right\|$ and $\left\|\bm g_{me}^k\right\|$. Specifically, when $\left\|\bm g_{mb}^k\right\|>\left\|\bm g_{me}^k\right\|$, in the high $P_m$ regime, i.e., high SNR case, $\phi_{m}^k$ can be approximated by
{\setlength\arraycolsep{2pt}% 用于缩短等号两边的空格
\begin{eqnarray}
\phi_{m}^k&\simeq& \frac{V}{K}\left(\log_2\frac{ P_m \left\|\bm g_{mb}^k\right\|^2}{N_0}- \log_2\frac{P_m \left\|\bm g_{me}^k\right\|^2}{N_0}\right), \nonumber\\
&=&\frac{2V}{K} \log_2 \frac{\left\|\bm g_{mb}^k\right\|}{\left\|\bm g_{me}^k\right\|},
\end{eqnarray}}
\!\!\!which is a constant, and implies that $\phi_{m}^k$ is asymptotically invariant as $P_m$ grows large. Thus, Lemma \ref{lemma1} is proven.
\section{Proof of Theorem \ref{theorem_3}}
\label{Appendix_C}
For notational brevity, denote
{\setlength\arraycolsep{2pt}% 用于缩短等号两边的空格
\begin{eqnarray}
&&\epsilon_m=\left|\sqrt{P_m} \bm w_m^H \bm g_{mb}^{k_m}-1\right|^2+N_0 \left\|\bm w_m \right\|^2, \nonumber\\
&&\delta_m^n=\left|\bm w_m^H \bm h_{nb}^{k_m}\right|^2, \,\varepsilon_m=P_m \left\|\bm g_{me}^{k_m}\right\|^4, \nonumber\\
&&\zeta_m^n=\left|\left(\bm g_{me}^{k_m}\right)^H \bm h_{ne}^{k_m}\right|^2, \,\kappa_m=N_0 \left\|\bm g_{me}^{k_m}\right\|^2.
\label{delta_epsilon}
\end{eqnarray}}
\!\!Then, ${\tilde \chi}_{m,n}^{k_m}$ in (\ref{secrecy_capacity5}) can be rewritten as
\begin{equation}
{\tilde \chi}_{m,n}^{k_m} = -\log_2 \left(q_n \delta_m^n+\epsilon_m\right) - \log_2\left(1+\frac{ \varepsilon_m}{q_n \zeta_m^n+\kappa_m}\right).
\end{equation}
Since $q_n \in [0, Q_n]$, there exist extreme points when maximizing ${\tilde \chi}_{m,n}^{k_m}$ with respect to (w.r.t.) $q_n$. By calculating the first-order derivative of ${\tilde \chi}_{m,n}^{k_m}$ and setting it to be 0, we have
{\setlength\arraycolsep{2pt}% 用于缩短等号两边的空格
\begin{eqnarray}
\!\!\!\!&&\frac{\partial {\tilde \chi}_{m,n}^{k_m}}{\partial q_n} = \nonumber\\
\!\!\!\!&&\frac{-q_n^2\delta_m^n\!\left(\zeta_m^n\right)^2\!-\!2 q_n\delta_m^n \zeta_m^n \kappa_m \!+\! \epsilon_m \varepsilon_m \zeta_m^n \!-\! \delta_m^n \varepsilon_m \kappa_m \!-\! \delta_m^n \kappa_m^2}{\left(q_n \delta_m^n+\epsilon_m\right)\left(q_n \zeta_m^n+\varepsilon_m+\kappa_m\right)\left(q_n \zeta_m^n+\kappa_m\right)\ln2}\nonumber\\
\!\!\!\!&&=0,
\label{derivative}
\end{eqnarray}}
\!\!\!which is a quadratic equation and has the discriminant
\begin{equation}
\Delta_m^n = 4\delta_m^n \varepsilon_m \left(\zeta_m^n\right)^2\left(\epsilon_m \zeta_m^n -\delta_m^n \kappa_m\right).
\label{discriminant}
\end{equation}
Since the parameters in (\ref{delta_epsilon}) are all nonnegative, when $\Delta_m^n\leq 0$, i.e., $\epsilon_m \zeta_m^n \leq \delta_m^n \kappa_m$, $\frac{\partial {\tilde \chi}_{m,n}^{k_m}}{\partial q_n}\leq 0$, indicating that $ {\tilde \chi}_{m,n}^{k_m}$ decreases with $q_n$ and $q_n^*=0$. When $\epsilon_m \zeta_m^n > \delta_m^n \kappa_m$, the two zero points of (\ref{derivative}) can be written as
\begin{equation}
{\bar a}_m^n = -\frac{\kappa_m}{\zeta_m^n}-\frac{\sqrt{\Delta_m^n}}{2\delta_m^n \left(\zeta_m^n\right)^2}, \,{\tilde a}_m^n = -\frac{\kappa_m}{\zeta_m^n}+\frac{\sqrt{\Delta_m^n}}{2\delta_m^n \left(\zeta_m^n\right)^2}.
\label{zero_points}
\end{equation}
Obviously, ${\bar a}_m^n$ is negative. As a result, the monotonicity of $ {\tilde \chi}_{m,n}^{k_m}$ w.r.t. $q_n$ depends on the relationship between ${\tilde a}_m^n$, $0$ and $Q_n$. When ${\tilde a}_m^n\leq 0$, $ {\tilde \chi}_{m,n}^{k_m}$ decreases monotonically with $q_n \in [0, Q_n]$, so $q_n^*=0$. When $0 < {\tilde a}_m^n\leq Q_n$, $ {\tilde \chi}_{m,n}^{k_m}$ first increases on interval $[0, {\tilde a}_m^n]$ and then decreases on interval $[{\tilde a}_m^n,Q_n]$, hence, $q_n^*={\tilde a}_m^n$. Otherwise, if ${\tilde a}_m^n\geq Q_n$, $ {\tilde \chi}_{m,n}^{k_m}$ increases monotonically with $q_n \in [0, Q_n]$, so $q_n^*=Q_n$. Thus, Theorem \ref{theorem_3} is proven.

\section*{Acknowledgments}
This work was supported by the NSFC (Nos. 61372106, 61471114, \& 61221002), NSTMP under 2016ZX03001016-003, the Six Talent Peaks project in Jiangsu Province under GDZB-005, Science and Technology Project of Guangdong Province under Grant 2014B010119001, the Scholarship from the China Scholarship Council (No. 201606090039), Program Sponsored for Scientific Innovation Research of College Graduate in Jiangsu Province under Grant KYLX16\_0221, and the Scientific Research Foundation of Graduate School of Southeast University under Grant YBJJ1651.

\bibliographystyle{IEEEtran}
\bibliography{IEEEabrv,MMM}
%\begin{thebibliography}
%\bibitem{BIB17}
%Simulating the SUI Channel Models, IEEE Standard 802.16.3c-01/53, Sep. 2001,http://www.ieee802.org/16/tg3/contrib/802163c-01_53.pdf.
%\end{thebibliography}

% that's all folks
\end{document}